\documentclass{aims}
\usepackage{amsmath}
  \usepackage{paralist,tikz}
  \usepackage{graphics} 
  \usepackage{epsfig,eepic} 
\usepackage{graphicx}  \usepackage{epstopdf}
 \usepackage[colorlinks=true]{hyperref}
\hypersetup{urlcolor=blue, citecolor=red}

\def\currentvolume{11}
 \def\currentissue{2}
  \def\currentyear{2019}
   \def\currentmonth{June}
    \def\ppages{X--XX}
     \def\DOI{10.3934/jgm.2019009}

\def\bbbe\mathbb{E}

\def\bbbc{{\mathbb C}}
\def\bbbe{\mathbb{E}}

\newtheorem{remark}{Remark}

\def\diag{\mbox{diag}\,}

\def\Conf{{\rm Conf}\,}
\def\im{{\rm Im}\,}

\def\Aut{\mbox{Aut}\,}
\def\fr#1{{\mathfrak{#1}}}
\def\openone{\leavevmode\hbox{\small1\kern-3.3pt\normalsize1}}

\numberwithin{equation}{section}

\def\im{\mbox{Im\,}}

\allowdisplaybreaks

\newtheorem{corollary}{Corollary}

\theoremstyle{definition}

\title[RHP, Integrability and reductions] 
      {Riemann-Hilbert problem, integrability\\ and reductions}

\author[Vladimir S. Gerdjikov,  Rossen I. Ivanov and Aleksander A. Stefanov]{}


 \email{vgerdjikov@gmail.com}
 \email{rossen.ivanov@dit.ie}
 \email{aleksander.a.stefanov@gmail.com}

\thanks{$^*$ Corresponding author: R. I. Ivanov}

\begin{document}
\maketitle


\centerline{\scshape Vladimir S. Gerdjikov$^{1,2,5}$,  Rossen I. Ivanov$^{3,*}$}
\centerline{\scshape  and  Aleksander A. Stefanov$^{2,4}$}
\medskip
{\footnotesize
 \centerline{$^1$ Department of Applied Mathematics}
   \centerline{National Research Nuclear University MEPHI}
      \centerline{ 31 Kashirskoe Shosse, Moscow 115409, Russian Federation}
   \centerline{$^2$ Institute of Mathematics and Informatics }
   \centerline{ Bulgarian Academy of Sciences }
\centerline{8 Acad. G. Bonchev Street, Sofia 1113, Bulgaria}
\centerline{$^3$ School of Mathematical Sciences}
\centerline{ Technological University Dublin - City Campus }
   \centerline{Kevin Street, Dublin D08 NF82, Ireland}
    \centerline{$^4$  Faculty of Mathematics and Infromatics}
   \centerline{Sofia University St. Kliment Ohridsky}
   \centerline{ 5 James Bourchier Blvd., Sofia 1164, Bulgaria}
\centerline{$^5$ Institute for Advanced Physical Studies}
   \centerline{   New Bulgarian University }
   \centerline{  21 Montevideo Street, Sofia 1618, Bulgaria}

} 

%

\begin{abstract}
The present paper is dedicated to integrable models with Mikhailov reduction groups $G_R \simeq \mathbb{D}_h.$ Their Lax representation allows us to prove, that their solution is equivalent to solving
Riemann-Hilbert problems, whose contours  depend on the realization of the $G_R$-action on the spectral parameter.
Two new examples of Nonlinear Evolution Equations (NLEE) with $\mathbb{D}_h$ symmetries are presented.
\end{abstract}

\section{Introduction}

One of the standard methods in the study of integrable systems is based on the Lax representation and the inverse scattering method \cite{ZMNP,FaTa,AblPrin,GVY}.
The Inverse Scattering Problem (ISP) for a given Lax operators requires the knowledge of the so-called Fundamental Analytic Solutions (FAS) \cite{Sha*74}. The next step is the transformation of the ISP to an equivalent Riemann-Hilbert problem (RHP) \cite{ZaSh*74a,ZaSh*79}, which proved to be very fruitful.

Indeed, already the scalar RHP provides a tool for the implementation of the inverse scattering method and for calculation of the soliton solutions of some very important and interesting equations such as
the KdV \cite{ZMNP}, the Kaup-Boussinesq \cite{IL,HL}, the Camassa-Holm \cite{CGI,HoI1} and the $2$-component Camassa-Holm \footnote{The Camassa-Holm equation introduced in \cite{CH93} triggered an enormous flow of publications and many interesting generalizations.}  \cite{HoI2} equations.

The more general problem is a RHP in matrix form
\begin{equation}\label{eq:rhp1}\begin{split}
\xi^+ (x,t,\lambda) =  \xi^- (x,t,\lambda)  G(x,t,\lambda), \qquad \lambda \in \mathbb{R}.
\end{split}\end{equation}
for Lie group-valued functions $\xi^\pm (x,t,\lambda)$,  analytic for $\lambda \in \mathbb{C}_\pm$.
Such RHP is a tool for solving the ISP   for the generalized Zakharov-Shabat system:
\begin{equation}\label{eq:gZS}\begin{split}
L_{\rm gZS} \psi \equiv i \frac{\partial \psi}{ \partial x } + (Q(x,t) - \lambda J) \psi(x,t,\lambda) =0,
\end{split}\end{equation}
where $J =\diag (a_1,\dots, a_n)$ is a diagonal matrix with different real-valued eigenvalues \cite{Sha*74,Sha*79}. Using $L_{\rm gZS}$
Zakharov and Manakov discovered the integrability of the $N$-wave equations \cite{ZaMan}, see also \cite{Kaup}.
Particular cases of $L_{\rm gZS}$, where some of the $a_j$ are equal, can be applied to the Manakov model \cite{Man} as well as to other vector (multi-component) nonlinear Schr\"odinger equations (NLS)
related to symmetric spaces.
Using RHP's  (\ref{eq:rhp1}) Zakharov and Shabat introduced their dressing method for calculating the soliton
solutions of all these Nonlinear Evolution Equations (NLEE) \cite{ZaSh*74a,ZaSh*79}.

Another important method for deriving new NLEE was proposed by Mikhailov \cite{Mikh}, where he introduced
the group of reductions $G_R$. The relevant Lax operator is of the form $L_{\rm gZS}$ but now $a_j$ are powers of
$\omega =\exp(2\pi i/h)$.

The problem of constructing FAS for generalized Zakharov-Shabat systems with complex-valued $a_j$
was solved by Beals and Coifman \cite{Beals-Coifman} for the $sl(N)$-algebras and was later generalized
to any simple Lie algebras \cite{GeYa*94}. Thus the ISP for Lax operators having $\mathbb{Z}_h$ and $\mathbb{D}_h$ as
reduction groups again can be reduced to a RHP, but now its contour in the $\lambda$-plane is a set of $h$ lines closing
angles $\pi/h$ (see fig. \ref{fig:1}):
\begin{equation}\label{eq:rhp1k}\begin{split}
\xi_k (x,t,\lambda) =  \xi_{k-1} (x,t,\lambda)  G_k(x,t,\lambda), \qquad \lambda \in l_k
, \quad k=0,\dots, h-1.
\end{split}\end{equation}
Here $\xi_k (x,t,\lambda) $ is an analytic function in the sector $k\pi/h \leq \arg \lambda \leq
(k+1)\pi/h$ and $k-1$ is evaluated modulo $h$. Among the relevant new NLEE
with $\mathbb{Z}_h$ and $\mathbb{D}_h$ symmetries related to
such RHP we note the 2-dimensional Toda field theories \cite{Mikh}, mKdV equations related to the
Kac-Moody algebras \cite{Za1*76,MaZa*79,Za*80,MiZa*80,DriSok,Za*80b}, see also \cite{ContM,GuKV,GGMV,107s,120a,SIAM,127a,129a} and others.

Already at this level we see that the formulation of the RHP depends substantially on the group
of reductions $G_R$ imposed on the Lax pair. Indeed, the contour $\Gamma$ of the RHP depends on the
explicit realization of $G_R$ as a subgroup of the group of conformal transformation $\Conf $ on
$\mathbb{C}$. From now on for definiteness we will assume that the reduction group $G_R \simeq \mathbb{D}_h$ is isomorphic to the dihedral group.

The 2-dimensional Toda field theories \cite{Mikh,MiOlPer} and the mKdV equations
\cite{DriSok} provide paradigmatic examples of NLEE with $\mathbb{Z}_h$ and $\mathbb{D}_h$ symmetries.
Their Lax representations in fact are closely related to the Kac-Moody algebras \cite{DriSok}.
The spectral properties of their Lax operators \cite{GeYa*13,113s,SIAM} however are not completely established and deserve further studies.

The main focus of this paper is on the Lax operators with a reduction group
$G_R  $ isomorphic to the dihedral group $\mathbb{D}_h$. We are especially interested in reductions containing the transformation
$\lambda \to \frac{1}{\lambda}$, where $\lambda$ is the spectral parameter. As a starting point though,
we will review the operators with a $\mathbb{Z}_h$ reduction group.

The paper is structured as follows. Section 2 contains some preliminaries. There we formulate Mikhailov's reduction group
(Subsection 2.1) as well as  the standard formulation of the RHP for Lax pairs with a $\mathbb{Z}_h$ reduction group.
In Section 3 we analyze Lax pairs with a $\mathbb{D}_h$ reduction groups.  We demonstrate that they are equivalent to
Riemann-Hilbert problems, whose contours  depend on the realization of the $G_R$-action on the spectral parameter.
Section 4 contains two new examples NLEE with $\mathbb{D}_h$ symmetries.
\textcolor{black}{The first one is a generalization of the  GI eq. It is in the list
of integrable NLS-type equations classified by Mikhailov, Shabat, Yamilov and Sokolov \cite{MiShaYa2,MiShaYa1,51a}. The second example
is a deformation of the Heisenberg ferromagnet equation.}

\section{Preliminaries}
Let us start with an example for a typical Lax pair, given below.
It is polynomial in the spectral parameter $\lambda$ and its potentials
$U(x,t,\lambda)$ and $V(x,t,\lambda)$ are taking values in the simple Lie algebra $\mathfrak{g}$.
\begin{equation}\label{eq:W1aN}\begin{split}
L\psi &\equiv i\frac{\partial \psi}{ \partial x } + U(x,t,\lambda) \psi (x,t,\lambda) =0, \quad
\\
M\psi &\equiv i\frac{\partial \psi}{ \partial t } + V(x,t,\lambda) \psi (x,t,\lambda) =0. \quad
\end{split}\end{equation}
Here $U(x,t,\lambda)$ and $V(x,t,\lambda)$ depend polynomially on $\lambda$ and with proper Mikhailov
reduction it can generate various important equations, like the mKdV and NLS-type equations related to $\mathfrak{g}$, see for example \cite{DriSok,GVY,ContM,GVY,107s,GI1,GI2,120a,R}.

\subsection{Mikhailov's reduction group}

The reduction group $G_R $ is a finite group which preserves the
Lax representation (\ref{eq:W1aN}), i.e. it ensures that the reduction
constraints are automatically compatible with the evolution. $G_R $ must
have two realizations: i) $G_R \subset {\rm Aut}\fr{g} $ and ii) $G_R
\subset {\rm Conf}\, \Bbb C $, i.e. as conformal mappings of the complex
$\lambda $-plane. To each $g_k\in G_R $ we relate a reduction
condition for the Lax pair as follows \cite{Mikh}:
\begin{equation}\label{eq:2.1}
C_k(L(\Gamma _k(\lambda ))) = \eta _k L(\lambda ), \quad
C_k(M(\Gamma _k(\lambda ))) = \eta _k M(\lambda ),
\end{equation}
where $C_k\in \mbox{Aut}\; \fr{g} $ and $\Gamma _k(\lambda )\in
\mbox{Conf\,} \bbbc $ are the images of $g_k $ and $\eta _k =1 $ or $-1 $
depending on the choice of $C_k $. Since $G_R $ is a finite group then for
each $g_k $ there exist an integer $N_k $ such that $g_k^{N_k} =\openone
$.

More specifically the automorphisms $C_k $, $k=1,\dots,4 $ listed above
lead to the following reductions for the matrix-valued functions
of the Lax representation:
\begin{equation}\label{eq:U-V}\begin{aligned}
&\mbox{1)} & \; C_1(U^{\dagger}(\kappa _1(\lambda )))&= U(\lambda ),
&\quad C_1(V^{\dagger}(\kappa _1(\lambda ))) &= V(\lambda ), \\
&\mbox{2)} & \; C_2(U^{T}(\kappa _2(\lambda )))&= -U(\lambda ), &\quad
C_2(V^{T}(\kappa _2(\lambda )))&= -V(\lambda ), \\
&\mbox{3)} & \;  C_3(U^{*}(\kappa _1(\lambda )))&= -U(\lambda ), &\quad
C_3(V^{*}(\kappa _1(\lambda )))&= -V(\lambda ), \\
&\mbox{4)} &\; C_4(U(\kappa _2(\lambda )))&= U(\lambda ), &\quad
C_4(V(\kappa _2(\lambda )))&= V(\lambda ),
\end{aligned}\end{equation}
 reductions of type 1) and 2) in (\ref{eq:U-V}) involve combinations
with outer automorphisms of $\mathfrak{g}$.

Thus the explicit form of the Lax representation and consequently of the NLEE, depends on:
(a) the choice of the reduction group $G_r$; (b) the choice of its representation as a subgroup
of the conformal group acting on $\lambda$; and (c) the choice of the representation of $G_R$ as a subgroup
of $\Aut \mathfrak{g}$. It is important also that both representations of $G_R$ are faithful.

\subsection{Lax pairs with $\mathbb{Z}_h$ symmetries. Single point orbits.} 

The cyclic group $\mathbb{Z}_h$ has one generating element satisfying the generating relation:
\begin{equation}\label{eq:Dh1}\begin{split}
s^h = \openone.
\end{split}\end{equation}
The group has $h$ elements: $\openone, s^k, k=1,\dots , h-1 $. Typically the action of $\mathbb{Z}_h$ is
\begin{equation}\label{eq:Dh2}\begin{split}
s(\lambda) = \lambda \omega,
\end{split}\end{equation}
where $\omega = \exp(2\pi i/h)$. Next, following \cite{Mikh,DriSok} we choose the Lax pair to be polynomial in $\lambda$. Furthermore, we assume that
\begin{equation}\label{eq:W1a}\begin{split}
 U(x,t,\lambda)&=Q(x,t) - \lambda J, \\
V(x,t,\lambda) &=\sum_{k=0}^{s-1}\lambda^k V_k(x,t)  - \lambda^s K,
\end{split}\end{equation}
where $\lambda$ is the spectral parameter and $U(x,t,\lambda)$ and $V(x,t,\lambda)$ take values in the simple Lie
algebra $\mathfrak{g}$. In other words we are using a degenerate orbit
of the $\mathbb{Z}_h$ group consisting of the only singular point $\{ \infty \} $ of the Lax pair.

Typically in what follows by $h$ we will denote the Coxeter number of $\mathfrak{g}$ and the corresponding
inner automorphism $C$ such that $C^h=\openone $ will be the Coxeter automorphism (either belonging to the
Weyl group or an equivalent to it automorphism from the Cartan subgroup).

The Coxeter automorphism introduces grading in the algebra $\mathfrak{g}$ \cite{Helgasson}, namely:
\begin{equation}\label{eq:Zhgra}\begin{split}
\mathfrak{g} = \mathop{\oplus}\limits_{s=0}^{h-1} \mathfrak{g}^{(s)}, \qquad
[\mathfrak{g}^{(s)}, \mathfrak{g}^{(p)}] \in \mathfrak{g}^{(s+p)},
\end{split}\end{equation}
with the properties
\begin{equation}\label{eq:Zhgra2}\begin{split}
X \in \mathfrak{g}^{(s)}  \quad \Leftrightarrow \quad C X C^{-1} = \omega^s X, \quad
Y \in \mathfrak{g}^{(p)}  \quad \Leftrightarrow \quad C YC^{-1} = \omega^p X,
\end{split}\end{equation}
i.e. $[X,Y] \in \mathfrak{g}^{(p+s)}$ and $p+s$ is understood modulo $h$.
In particular, the Lax operator $L$ in (\ref{eq:W1a}) will possess $\mathbb{Z}_h$ symmetry
if $Q(x,t) \in \mathfrak{g}^{(0)}$ and $J \in \mathfrak{g}^{(1)}$.

The spectral properties of such Lax operators are well known, see \cite{GeYa*94,SIAM} and
the references therein. The continuous spectrum of $L$ fills up the set of rays
\begin{equation}\label{eq:Gam}\begin{split}
\Gamma = \mathop{\oplus}\limits_{s=0}^{2h-1} l_s, \qquad l_s \equiv\left \{ \lambda \quad \arg \lambda = \frac{s \pi}{h}, \;
k=0, \dots, 2h-1 \right \}.
\end{split}\end{equation}
These rays split the complex plane $\mathbb{C}$ into $2h$ sectors $\Omega_k : k \pi/h \leq \arg \lambda
\leq (k+1) \pi/h$, $k=0,\dots, 2h-1$, see figure \ref{fig:1}.

Let $\Delta$ be the root system of $\mathfrak{g}$. As usual, it splits into two parts $\Delta = \Delta_+ \cup \Delta_-$,
where $\Delta_+$ denotes the set of the positive roots, and $\Delta_-$ the set of the negative roots. Then  to each ray we can relate a subset of roots
\begin{equation}\label{eq:delk}\begin{split}
\delta_k \equiv \{ \alpha \in \Delta, \qquad \im \lambda \alpha(J) =0, \quad \lambda \in l_k \}.
\end{split}\end{equation}
Each of the subsets also can be split into positive and negative roots by
\begin{equation}\label{eq:delk2}\begin{split}
\delta_k^+ &\equiv \{ \alpha \in \delta_k, \qquad \im \lambda \alpha(J) >0, \quad \lambda \in e^{i0} l_k \}, \\
\delta_k^- &\equiv \{ \alpha \in \delta_k, \qquad \im \lambda \alpha(J) <0, \quad \lambda \in e^{-i0} l_k \}.
\end{split}\end{equation}
Note, that this ordering is specific for each ray $l_k$.

With each of the sectors $\Omega_k$ one can relate a FAS of $L$ $\chi_k(x,t,\lambda)$.
Next we can introduce the limits of the FAS for $x\to \pm \infty$ which will play the role of
the minimal sets of scattering data of $L$, see \cite{GeYa*94,SIAM}:
\begin{equation}\label{eq:Skpm}\begin{split}
\lim _{x\to -\infty} e^{-i\lambda J x} \chi_k(x,t,\lambda) = S_k^\pm, \quad
\lim _{x\to  \infty} e^{-i\lambda J x} \chi_k(x,t,\lambda) = T_k^\mp D_k^\pm, \quad \lambda \in e^{\pm i0} l_k ,
\end{split}\end{equation}
where
\begin{equation}\label{eq:Skpm_exp}\begin{aligned}
S_k^\pm (\lambda,t) &= \exp \left( \sum_{\alpha \in \delta_k^+}^{} s^\pm_\alpha (\lambda,t) E_{\pm \alpha} \right), &\;
D_k^\pm (\lambda) &= \exp \left( \sum_{\alpha \in \delta_k^+}^{} d^\pm_\alpha (\lambda) H_{\alpha} \right), \\
T_k^\pm (\lambda,t) &= \exp \left( \sum_{\alpha \in \delta_k^+}^{} \tau^\pm_\alpha (\lambda,t) E_{\pm \alpha} \right),
\end{aligned}\end{equation}

As minimal sets of scattering data one can use each of the sets of functions: (i) $S_0^\pm (\lambda,t)$ and $S_1^\pm (\lambda,t) $
for $\lambda \in l_0$ and $\lambda \in l_1$ respectively; or (ii) $T_0^\pm (\lambda,t)$ and $T_1^\pm (\lambda,t) $
for $\lambda \in l_0$ and $\lambda \in l_1$ respectively, \cite{GeYa*94,SIAM}. Note that all other sets of functions
$S_k^\pm (\lambda,t)$ (resp. $T_k^\pm (\lambda,t)$) are recovered from $S_0^\pm (\lambda,t)$ and $S_1^\pm (\lambda,t) $
(resp. $T_0^\pm (\lambda,t)$ and $T_1^\pm (\lambda,t) $) by the symmetry conditions:
\begin{equation}\label{eq:Spmk}\begin{split}
 S_{2s+k}^\pm (\lambda \omega^s,t) = C^s S_k^\pm (\lambda,t)C^{-s}, \qquad
T_{2s+k}^\pm (\lambda \omega^s,t) = C^s T_k^\pm (\lambda,t)C^{-s},
\end{split}\end{equation}
with $ k=0,1$.

From eqs. (\ref{eq:Skpm}) it follows, that the FAS satisfy:
\begin{equation}\label{eq:rhp0}\begin{split}
\chi_{k} (x,t,\lambda) = \chi_{k-1} (x,t,\lambda) G_{k}^{(0)} (t,\lambda) , \quad
G_{k}^{(0)} (t,\lambda)= S_{k}^{-, -1} (t,\lambda) S_{k}^+ (t,\lambda),
\end{split}\end{equation}
for $k=0, \dots, 2h-1$. In order to derive the proper RHP we need to introduce:
\begin{equation}\label{eq:rhp1kc}\begin{split}
\xi_{k} (x,t,\lambda) = \chi_{k} (x,t,\lambda) e^{iJ\lambda x}.
\end{split}\end{equation}
Then the relations (\ref{eq:rhp0}) can be viewed as a RHP on the contour $\Gamma$ allowing canonical normalization:
\begin{equation}\label{eq:rhp1kd}\begin{split}
&\xi_{k} (x,t,\lambda) = \xi_{k-1} (x,t,\lambda) G_{k} (x,t,\lambda) , \\
&\lim_{\lambda \to \infty} \xi_k(x,t,\lambda) = \openone.
\end{split}\end{equation}
The $x$-dependence of the sewing function is given by
\begin{equation}\label{eq:rhpGk}\begin{split}
G_{k} (x,t,\lambda) &= e^{i\lambda Jx} G_{k}^{(0)} (t,\lambda) e^{-i\lambda Jx}, \quad \lambda \in l_k.
\end{split}\end{equation}
Thus, starting from the Lax operator and its FAS we derived canonically normalized RHP.

An important fact, proved by Zakharov and Shabat \cite{ZaSh*74a,ZaSh*79} is, that from the RHP (\ref{eq:rhp1kc})
one can derive the Lax operator. Indeed, following Zakharov and Shabat \cite{ZaSh*74a,ZaSh*79} it became possible to
develop the dressing method for explicit calculation of the soliton solutions.
Later results by Dickey, Gelfand \cite{GelDi} and Mark Adler \cite{MAdler} established the fact that
the solution of the RHP in each of the sectors $\Omega_k$ for $\lambda \gg 1$ can be represented as an asymptotic
series in the form \cite{Pliska}:
\begin{equation}\label{eq:rhp3}\begin{split}
\xi_k (x,t,\lambda) = \exp (\mathcal{Q}^+ (x,t,\lambda)), \quad
\mathcal{Q}^+ (x,t,\lambda) = \sum_{s=1}^{\infty} Q_s(x,t) \lambda^{-s}, \quad \lambda \gg 1.
\end{split}\end{equation}
This representation is compatible with the canonical normalization of RHP and allows to parameterise
the Lax pair by:
\begin{equation}\label{eq:LaxUV}\begin{split}
U^+(x,t,\lambda) = - (\lambda \xi_k J \xi_k^{-1})_+, \qquad
V^+(x,t,\lambda) = - (\lambda^3 \xi_k J \xi_k^{-1})_+,
\end{split}\end{equation}
where the subscript $+$ means that only terms with positive powers of $\lambda$ in the corresponding expansion in powers of $\lambda$ are retained. In \cite{Pliska} it was demonstrated that the first
few coefficients of $\mathcal{Q}(x,t,\lambda)$ were sufficient to parameterize $U$ and $V$ of a polynomial in $\lambda$ Lax pair. Indeed, for a Lax operator linear in $\lambda$ we have
\begin{equation}\label{eq:Up}\begin{split}
U^+(x,t,\lambda) = - (\lambda \xi_k J \xi_k^{-1})_+ = - \lambda J + [J, Q_1(x,t)].
\end{split}\end{equation}

The formulae (\ref{eq:rhp3}) and (\ref{eq:LaxUV}) allow one to calculate effectively compatible
Lax pairs when both $L$ and $M$ are quadratic in $\lambda$ or polynomials in $\lambda$ of any order greater than $2$,
see \cite{Pliska}. One can also impose on $\mathcal{Q}^+(x,t,\lambda)$ reductions of $\mathbb{Z}_h$-type; the resulting
Lax pair will naturally possess $\mathbb{Z}_h$ as reduction group.

\section{Lax pairs with $\mathbb{D}_h$ symmetries}

\subsection{$\mathbb{D}_h$ symmetries: General construction.}
The dihedral group $\mathbb{D}_h$ has two generating elements satisfying the generating relations:
\begin{equation}\label{eq:Dh11}\begin{split}
r^2 = s^h = \openone, \qquad srs^{-1} = s^{-1} .
\end{split}\end{equation}
The group has $2h$ elements: $\{ s^k, rs^k, k=1,\dots , h\} $ and allows several inequivalent
realization on the complex $\lambda$-plane. Some of them are:
\begin{equation}\label{eq:Dh2a}\begin{aligned}
& \mbox{(i)} &\;  s(\lambda) &= \lambda \omega, &\; r(\lambda) &= \epsilon \lambda^* , &\quad
& \mbox{(ii)} &\;  s(\lambda) &= \lambda \omega, &\; r(\lambda) &= \frac{\epsilon}{\lambda^*} ,\\
& \mbox{(iii)} &\;  s(\lambda) &= \lambda \omega, &\; r(\lambda) &= \epsilon \lambda , &\quad
& \mbox{(iv)} &\;  s(\lambda) &= \lambda \omega, &\; r(\lambda) &= \frac{\epsilon}{\lambda} ,
\end{aligned}\end{equation}
where $\omega = \exp(2\pi i/h)$ and $\epsilon =\pm 1$.
In Section 4 we will present two examples of new NLEE whose Lax representation
conform with case (iv) of (\ref{eq:Dh2a}).

Detailed analysis of the realizations of $\mathbb{D}_h$ as reduction groups of Lax pairs has
been performed by Mikhailov, Lombardo and Bury \cite{LoMikh1,LoMikh2,KniLoSa,LoSa,Rhys,Mik3}. It depends substantially
not only on the realization of $r(\lambda)$, but also on the choice of the orbit of $\mathbb{D}_h$.
Here we will limit ourselves to the two-point orbit containing the points $\{ 0, \infty\}$.

We will also need a pair of automorphisms $C_s$ and $C_r$ such that
\begin{equation}\label{eq:Dh1b}\begin{split}
C_s^h = \openone, \qquad C_r^2 = \openone, \qquad C_s C_r C_s^{-1} = C_s^{-1}.
\end{split}\end{equation}

\begin{figure}

\begin{tikzpicture}
\draw (0,0) -- (4,0);
\draw (-4,0) -- (0,0);
\draw (0,0) -- (60:4);
\draw (0,0) -- (120:4);
\draw (0,0) -- (240:4);
\draw (0,0) -- (300:4);
\draw (0,0) -- (120:4);

\coordinate (L0) at (05:3.9);
\coordinate (L1) at (65:3.9);
\coordinate (L2) at (125:3.9);
\coordinate (L3) at (175:3.9);
\coordinate (L4) at (235:3.9);
\coordinate (L5) at (305:3.9);

\coordinate (P0) at (30:3);
\coordinate (P1) at (90:3);
\coordinate (P2) at (150:3);
\coordinate (P3) at (210:3);
\coordinate (P4) at (270:3);
\coordinate (P5) at (330:3);

\node at (P0)  {$\Omega_0$};
\node at (P1)  {$\Omega_1$};
\node at (P2)  {$\Omega_2$};
\node at (P3)  {$\Omega_3$};
\node at (P4)  {$\Omega_4$};
\node at (P5)  {$\Omega_5$};

\node at (L0)  {$l_0$};
\node at (L1)  {$l_1$};
\node at (L2)  {$l_2$};
\node at (L3)  {$l_3$};
\node at (L4)  {$l_4$};
\node at (L5)  {$l_5$};

\node at (3.9,3.9) [draw, shape=circle] {$\lambda$};
\end{tikzpicture}
  \caption{Contour of a RHP with $\mathbb{Z}_3$ symmetry}\label{fig:1}
\end{figure}
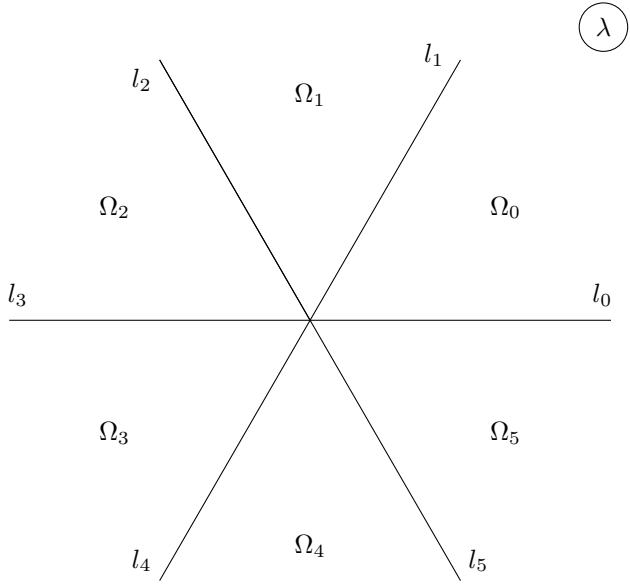

The realizations (i) and (iii) of $\mathbb{D}_h$  were used by Mikhailov to derive the 2-dimensional
Toda field theories \cite{Mikh} and by Drinfeld and Sokolov for obtaining the mKdV equations related
to the Kac-Moody algebras. The potentials of the relevant Lax representations take values in the simple
Lie algebras $\mathfrak{g}$ of rank $r$, which is graded using the Coxeter automorphism of order $h$.
Each of the above mentioned NLEE is a system of $r$ equations for $r$ real-valued functions.
The fact that the corresponding inverse scattering problem is equivalent to a RHP was noted by Mikhailov \cite{Mikh}.
It has the form (\ref{eq:rhp1kc}) and the contour is the union of $h$ rays, see fig. \ref{fig:1}. The additional $\mathbb{Z}_2$ generating element $r$  in the group  does not affect the contour of the RHP because
the FAS of the Lax operators are again analytic in the sectors $\Omega_k$, see fig. \ref{fig:1} where the case with $h=3$ is shown.

The situation changes if we consider the cases (ii) and (iv) in (\ref{eq:Dh2a}). Then the Lax pair becomes polynomial in $\lambda$ and  $\lambda^{-1}$, see  \cite{Rhys,LoMikh1}.
Now we must use two asymptotic expansions  of $\xi_k(x,t,\lambda)$: one for $|\lambda| \gg 1$
and another valid for  $|\lambda| \ll 1$:
\begin{equation}
\begin{split}
\xi_k^+ (x,t,\lambda) &= \exp (\mathcal{Q}^+ (x,t,\lambda)), \quad
\mathcal{Q}^+ (x,t,\lambda) = \sum_{s=1}^{\infty} Q_s(x,t) \lambda^{-s}, \quad \lambda \gg 1, \nonumber
\end{split}\end{equation}\begin{equation}\label{eq:rhp3b}\begin{split}\xi_{k+2h}^- (x,t,\lambda) &= \exp (\mathcal{Q}^- (x,t,\epsilon\lambda)), \quad
\mathcal{Q}^- (x,t,\lambda) = \sum_{s=1}^{\infty} C_r(Q)_s(x,t) (\epsilon\lambda)^{s}, \quad \lambda \ll 1.
\end{split}\end{equation}
Obviously the second asymptotic expansion in (\ref{eq:rhp3b}) is obtained from the first one by the action of the reduction group. In other words the solution of the RHP in each of the sectors $\Omega_{k+h}$ for $\lambda \ll 1$ can be obtained from $\xi_k^+(x,\lambda)$ by applying the automorphism $C_r$ : $\xi_k^-(x,\lambda)=C_r( \xi_k^+)(x,\epsilon\lambda^{-1})$.

Now equations (\ref{eq:LaxUV}) allow us to derive the parts of $U(x,t,\lambda)$ and $V(x,t,\lambda)$ that are polynomial
in $\lambda$. Analogously, using the formulae
\begin{equation}\label{eq:LaxUVm}\begin{split}
U^-(x,t,\lambda) = - (\lambda^{-1} \xi_k^- J \xi_k^{-,-1})_-, \qquad
V^-(x,t,\lambda) = - (\lambda^{-1} \xi_k^- J \xi_k^{-,-1})_-,
\end{split}\end{equation}
we derive the parts of $U(x,t,\lambda)$ and $V(x,t,\lambda)$ that are polynomial in $\lambda^{-1}$.
The subscript ``$-$'' in eq. (\ref{eq:LaxUVm}) means that only the non-positive powers of the expansion of the corresponding expressions in powers
of $\lambda$ is retained. The first few coefficients of $\mathcal{Q}(x,t,\lambda)$ are sufficient to parameterize $U^-$ and $V^-$ for
the non-polynomial in $\lambda$ part of the Lax pairs:
\begin{equation}\label{eq:U-}\begin{split}
U^-(x,t,\lambda) = - (\epsilon\lambda^{-1} \xi_k^- C_r(J) \xi_k^{-,-1})_- =
- \frac{1}{\epsilon\lambda} C_r(J) + [C_r(J), C_r(Q)_1(x,t)], \\
V^-(x,t,\lambda) = - (\epsilon\lambda^{-1} \xi_k^- C_r(K) \xi_k^{-,-1})_- =
- \frac{1}{\epsilon\lambda} C_r(K) + [C_r(K), C_r(Q)_1(x,t)].
\end{split}\end{equation}
Taking the average of these two procedures allows one to obtain a Lax representation, invariant with respect
to this representation of the dihedral group.

Important examples whose Lax pairs are with $\mathbb{Z}_3$-symmetry are the Degasperis-Procesi equation \footnote{The scalar Lax pair is derived in \cite{DHH}.} \cite{CIL}, the Tzitzeica equation \cite{BCG1,BCG2} and the Kaup-Kupershmidt equation \cite{Ka}.

Indeed, let us assume that the Lax pair $L$ and $M$ (\ref{eq:W1aN}) already possesses $\mathbb{Z}_h$ symmetry
with one point orbit. Using it we can construct new Lax pair $\tilde{L}$ and $\tilde{M}$ with $\mathbb{D}_h$ symmetry
with two-point orbit  as follows:
\begin{equation}\label{eq:LaxD2}\begin{split}
\tilde{ L} \psi &\equiv i\frac{\partial \psi}{ \partial x } + (U(x,t, \lambda) + C_r(U (x,t, \epsilon/\lambda)))
{\psi}(x,t,\lambda)=0, \\
\tilde{ M} \psi &\equiv i\frac{\partial \psi}{ \partial x } + (V(x,t,\lambda)+ C_r(V(x,t, \epsilon/\lambda)))
{ \psi}(x,t,\lambda)=0,
\end{split}\end{equation}

Now, depending on the specific choice of $C_r$  the FAS of $\tilde{L}$ (\ref{eq:LaxD2}) may have substantially different analytic properties. The contour of the RHP consists of $2h$-rays closing angles $\pi/h$ intersected by the unit circle $S^1$. As a result FAS can be defined in each of the $4h$ sectors
and the formulation of the RHP changes into:
\begin{equation}\label{eq:RHP2}\begin{aligned}
\xi_{k}^+ (x,t,\lambda) &= \xi_{k-1}^+ (x,t,\lambda) G_{k}^+ (x,t,\lambda) , &\quad \lambda &\in l_k^+, &\quad
k&=0,\dots, 2h-1 \\
\xi_{k}^- (x,t,\lambda) &= \xi_{k-1}^- (x,t,\lambda) G_{k}^- (x,t,\lambda) , &\quad \lambda &\in l_k^-, &\quad
k&=h,\dots, 2h-1 \\
\xi_{k+2h}^{(0)} (x,t,\lambda) &= \xi_{k}^{(0)} (x,t,\lambda) G_{k}^{(0)} (x,t,\lambda) , &\quad \lambda &\in a_k, &\quad
k&=0,\dots, 2h-1 ,
\end{aligned}\end{equation}
where by $l_k^+$ (resp. $l_k^-$) is the the part of the ray $l_k$ with $|\lambda|>1$ (resp. $|\lambda|<1$) and $a_k$ are the arcs
$k\pi/h \leq \arg \lambda \leq (k+1)\pi/h $ of the unit circle.

The $x$ dependence of the sewing functions $G_{k}^{(b)} (x,t,\lambda)$, $b=0,\pm$ is defined by:
\begin{equation}\label{eq:RHP2a}\begin{aligned}
i \frac{\partial  G_{k}^{\pm}}{ \partial x } - [ \mathcal{J}(\lambda), G_{k}^{\pm} (x,t,\lambda) ] &=0 , &\quad \lambda &\in l_k^\pm, &\quad
k&=0,\dots, h-1 , \\
i \frac{\partial  G_{k}^{(0)}}{ \partial x } - [ \mathcal{J}(\lambda), G_{k}^{(0)} (x,t,\lambda) ] &=0 , &\quad \lambda &\in a_k, &\quad
k&=0,\dots, h-1 ,
\end{aligned}\end{equation}
where $\mathcal{J}(\lambda)$ is invariant with respect to the reduction group $G_R$. Similarly the $t$ dependence will be given by:
\begin{equation}\label{eq:RHP2b}\begin{aligned}
i \frac{\partial  G_{k}^{\pm}}{ \partial t } - [ \mathcal{K}(\lambda), G_{k}^{\pm} (x,t,\lambda) ] &=0 , &\quad \lambda &\in l_k^\pm, &\quad
k&=0,\dots, h-1 , \\
i \frac{\partial  G_{k}^{(0)}}{ \partial t } - [ \mathcal{K}(\lambda), G_{k}^{(0)} (x,t,\lambda) ] &=0 , &\quad \lambda &\in a_k, &\quad
k&=0,\dots, h-1 ,
\end{aligned}\end{equation}
where $\mathcal{K}(\lambda)$ must also be invariant with respect to the reduction group $G_R$. The explicit form of $\mathcal{J}(\lambda)$ and $\mathcal{K}(\lambda)$ depends on the chosen model. The contours of the RHP for the case $h=3$ are given in fig. \ref{fig:2}.

\begin{figure}

\begin{tikzpicture}
\draw (0,0) -- (4,0);
\draw (-4,0) -- (0,0);
\draw (0,0) -- (60:4);
\draw (0,0) -- (120:4);
\draw (0,0) -- (240:4);
\draw (0,0) -- (300:4);
\draw (0,0) -- (120:4);
\draw (0,0) circle (2);

\coordinate (L0) at (05:3.9);
\coordinate (L1) at (65:3.9);
\coordinate (L2) at (125:3.9);
\coordinate (L3) at (175:3.9);
\coordinate (L4) at (235:3.9);
\coordinate (L5) at (305:3.9);

\coordinate (P0) at (30:3);
\coordinate (P1) at (90:3);
\coordinate (P2) at (150:3);
\coordinate (P3) at (210:3);
\coordinate (P4) at (270:3);
\coordinate (P5) at (330:3);

\coordinate (A0) at (30:2.2);
\coordinate (A1) at (90:2.2);
\coordinate (A2) at (150:2.2);
\coordinate (A3) at (210:2.2);
\coordinate (A4) at (270:2.2);
\coordinate (A5) at (330:2.2);

\coordinate (P6) at (30:1.5);
\coordinate (P7) at (90:1.5);
\coordinate (P8) at (150:1.5);
\coordinate (P9) at (210:1.5);
\coordinate (P10) at (270:1.5);
\coordinate (P11) at (330:1.5);

\node at (P0)  {$\Omega_0$};
\node at (P1)  {$\Omega_1$};
\node at (P2)  {$\Omega_2$};
\node at (P3)  {$\Omega_3$};
\node at (P4)  {$\Omega_4$};
\node at (P5)  {$\Omega_5$};

\node at (P6)  {$\Omega_6$};
\node at (P7)  {$\Omega_7$};
\node at (P8)  {$\Omega_8$};
\node at (P9)  {$\Omega_9$};
\node at (P10)  {$\Omega_{10}$};
\node at (P11)  {$\Omega_{11}$};

\node at (L0)  {$l_0$};
\node at (L1)  {$l_1$};
\node at (L2)  {$l_2$};
\node at (L3)  {$l_3$};
\node at (L4)  {$l_4$};
\node at (L5)  {$l_5$};

\node at (A0)  {$a_0$};
\node at (A1)  {$a_1$};
\node at (A2)  {$a_2$};
\node at (A3)  {$a_3$};
\node at (A4)  {$a_4$};
\node at (A5)  {$a_5$};

\node at (3.9,3.9) [draw, shape=circle] {$\lambda$};

\end{tikzpicture}

 \caption{Contour of the RHP $\mathbb{D}_3$ symmetry}\label{fig:2}
\end{figure}

\subsection{Equivalence of the RHP to a Lax representation}

One of the famous results of Zakharov and Shabat was the proof of the equivalence
between the RHP and the Lax representation \cite{ZaSh*74a,ZaSh*79}.

Our aim in this subsection is to outline the analog of Zakharov--Shabat theorem for
the case of a RHP with a $\mathbb{D}_h$ reduction group. Following the ideas of \cite{ZaSh*74a,ZaSh*79}
we introduce the functions
\begin{equation}\label{eq:g-k}\begin{split}
g_k^{b}(x,t) = i \frac{\partial \xi_k^{b}}{ \partial x } (\xi_k^{b})^{-1} - \xi_k^{b} \mathcal{J}(\lambda) (\xi_k^{b})^{-1}, \\
h_k^{b}(x,t) = i \frac{\partial \xi_k^{b}}{ \partial t } (\xi_k^{b})^{-1} - \xi_k^{b} \mathcal{K}(\lambda) (\xi_k^{b})^{-1},
\end{split}\end{equation}
with $b=0,\pm$. Then it is not difficult to show that
\begin{equation}\label{eq:gk}\begin{split}
g_k^{b}(x,t,\lambda) = g_{k-1}^{b}(x,t,\lambda) , \qquad h_k^{b}(x,t,\lambda) = h_{k-1}^{b}(x,t,\lambda) ,
\end{split}\end{equation}
for all $k=0,\dots , h-1$. The proof uses the conditions that both $\mathcal{J}(\lambda)$ and
$\mathcal{K}(\lambda)$ are invariant under the action of the group $\mathbb{D}_h$, and eqs. (\ref{eq:RHP2b}) hold.
The functions $g_{k}^{b}(x,t,\lambda)$ and $h_{k}^{b}(x,t,\lambda)$ satisfy the appropriate reduction conditions:
\begin{equation}\label{eq:RC1}\begin{split}
C_s\left( g_{k}^{b}(x,t,\lambda\omega)\right)  &= g_{k+1}^{b}(x,t,\lambda), \qquad
C_r \left(  g_{k}^{b}(x,t,\lambda)\right) = g_{k+2h}^{b} \left (x,t,\frac{\epsilon}{\lambda}\right ), \\
C_s\left(  h_{k}^{b}(x,t,\lambda\omega) \right) &= h_{k+1}^{b}(x,t,\lambda), \qquad
C_r \left( h_{k}^{b}(x,t,\lambda)\right)  = h_{k+2h}^{b}\left (x,t,\frac{\epsilon}{\lambda^*}\right ), \\
\end{split}\end{equation}
This means that  $g_{k}^{b}(x,t,\lambda)$ and $h_{k}^{b}(x,t,\lambda)$
take values in the automorphic  Lie algebra $\mathcal{A}(\mathfrak{g})$ introduced by
Mikhailov and Lombardo \cite{LoMikh1}. In our case (two-point orbit of $\mathbb{D}_h$) it will be enough to introduce basis in
the algebra $\mathcal{A}(\mathfrak{g})$ which is $\lambda$-dependent and is the analog of  Cartan-Weyl basis:
\begin{equation}\label{eq:Ea}\begin{split}
\mathcal{E}_\alpha^{(k)}(\lambda) = \lambda^k E_\alpha +  C_r(\lambda^{-k}{E}_{\alpha}), \qquad
\mathcal{H}_p^{(k)}(\lambda) = \lambda^k H_p +  C_r(\lambda^{-k}{H}_{p}).
\end{split}\end{equation}
By $C_r(\lambda^{-k}{E}_{\alpha})$ (resp. $C_r(\lambda^{-k}{E}_{\alpha})$) we mean  the action of the
reduction group element both on $\lambda$ and the Weyl generator $E_\alpha$ (resp. the Cartan generator $H_p$).
Obviously, also the potentials of the Lax operators will be taking values in $\mathcal{A}(\mathfrak{g})$.
One can also prove the following
\begin{corollary}\label{cor:1}
Let us consider the function $U(x,t,\lambda) \in \mathcal{A}(\mathfrak{g})$ such that for $\lambda\to\infty$
\begin{equation}\label{eq:Upm}\begin{split}
 \lim_{\lambda\to\infty} U(x,t,\lambda) \simeq \sum_{k=1}^{N_0} \lambda^k
 \left( \sum_{ \alpha}^{} U_{k,\alpha} (x,t) E_\alpha + \sum_{p=1}^{r} U_p(x,t) H_p \right) .
\end{split}\end{equation}
Then
\begin{equation}\label{eq:Upm1}\begin{split}
 U(x,t,\lambda) = \sum_{k=1}^{N_0} \left( \sum_{ \alpha}^{} U_{k,\alpha} (x,t)\mathcal{ E}_\alpha^{(k)}(\lambda) + \sum_{p=1}^{r}
U_p(x,t) \mathcal{ H}_p^{(k)}(\lambda) \right) .
\end{split}\end{equation}
\end{corollary}
Thus we conclude that the asymptotics of $U(x,t,\lambda)$ for $\lambda\to\infty$ determines it uniquely as a function taking values in $\mathcal{A}(\mathfrak{g}) $. The same holds true also for the functions
 $g_{k}^{b}(x,t,\lambda)$ and $h_{k}^{b}(x,t,\lambda)$.

\begin{remark}\label{rem:1}
\emph{The asymptotic of $g_{k}^{b}(x,t,\lambda)$ and $h_{k}^{b}(x,t,\lambda)$ for $\lambda \to \infty$
determine them uniquely. Indeed, we just need to combine this asymptotic with the  reduction conditions (\ref{eq:RC1}).}
\end{remark}

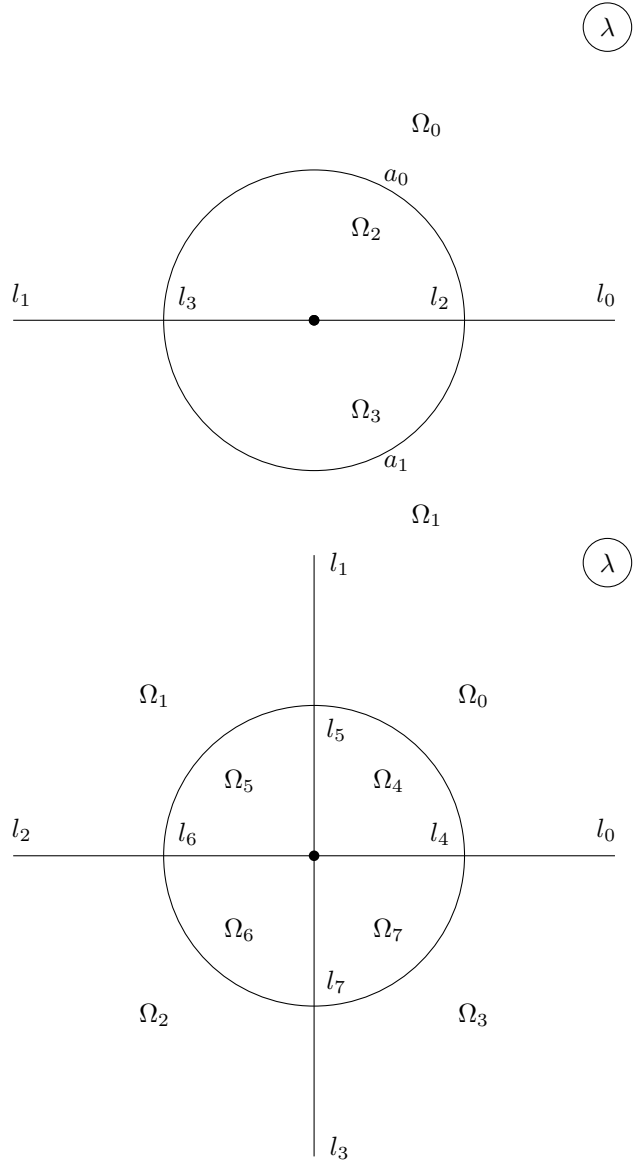
\begin{figure}

\begin{tikzpicture}
\fill(0,0) circle (2pt); \fill(0,0) circle (2pt);
\draw (0,0) -- (4,0);
\draw (-4,0) -- (0,0);
\draw (0,0) circle (2);
\coordinate (L0) at (05:3.9);
\coordinate (L3) at (175:3.9);
\coordinate (Ll0) at (10:1.7);
\coordinate (Ll3) at (170:1.7);
\coordinate (l0) at (60:1.4);
\coordinate (l3) at (300:1.4);
\coordinate (P0) at (60:3);
\coordinate (P5) at (300:3);
\coordinate (A0) at (60:2.2);
\coordinate (A5) at (300:2.2);
\coordinate (P6) at (60:1.5);
\coordinate (P10) at (300:1.5);

\node at (P0)  {$\Omega_0$};
\node at (P5)  {$\Omega_1$};
\node at (l0)  {$\Omega_2$};
\node at (l3)  {$\Omega_3$};
\node at (L0)  {$l_0$};
\node at (L3)  {$l_1$};
\node at (Ll0)  {$l_2$};
\node at (Ll3)  {$l_3$};
\node at (A0)  {$a_0$};
\node at (A5)  {$a_1$};
\node at (3.9,3.9) [draw, shape=circle] {$\lambda$};

\end{tikzpicture}

\begin{tikzpicture}
\fill(0,0) circle (2pt);
\draw (0,0) -- (4,0);
\draw (-4,0) -- (0,0);
\draw (0,0) -- (0,4);
\draw (0,-4) -- (0,0);
\draw (0,0) circle (2);

\coordinate (L0) at (05:3.9);
\coordinate (L1) at (85:3.9);
\coordinate (L2) at (175:3.9);
\coordinate (L3) at (275:3.9);
\coordinate (L4) at (10:1.7);
\coordinate (L5) at (80:1.7);
\coordinate (L6) at (170:1.7);
\coordinate (L7) at (280:1.7);

\coordinate (l4) at (45:1.4);
\coordinate (l5) at (135:1.4);
\coordinate (l6) at (225:1.4);
\coordinate (l7) at (315:1.4);

\coordinate (P0) at (45:3);
\coordinate (P1) at (135:3);
\coordinate (P2) at (225:3);
\coordinate (P3) at (315:3);

\coordinate (A0) at (60:2.2);
\coordinate (A5) at (300:2.2);

\coordinate (P6) at (60:1.5);
\coordinate (P10) at (300:1.5);

\node at (P0)  {$\Omega_0$};
\node at (P1)  {$\Omega_1$};
\node at (P2)  {$\Omega_2$};
\node at (P3)  {$\Omega_3$};

\node at (l4)  {$\Omega_4$};
\node at (l5)  {$\Omega_5$};
\node at (l6)  {$\Omega_6$};
\node at (l7)  {$\Omega_7$};

\node at (L0)  {$l_0$};
\node at (L1)  {$l_1$};
\node at (L2)  {$l_2$};
\node at (L3)  {$l_3$};
\node at (L4)  {$l_4$};
\node at (L5)  {$l_5$};
\node at (L6)  {$l_6$};
\node at (L7)  {$l_7$};

\node at (3.9,3.9) [draw, shape=circle] {$\lambda$};

\end{tikzpicture}

 \caption{Contour of the RHP for $\mathbb{D}_2$ symmetry
 (upper panel) and for $\mathbb{D}_4$ symmetry (lower panel)}\label{fig:3}
\end{figure}

\section{Examples}

\subsection{Generalization of the GI equation \cite{GI1,GI2}}
Consider the Lax pair
\begin{equation}\label{eq:zz}
\begin{aligned}
L\psi &\equiv i\frac{\partial \psi}{ \partial x } +U(x,t, \lambda) \psi (x,t,\lambda) =0, \\
M\psi &\equiv i\frac{\partial \psi}{ \partial t } +V(x,t, \lambda) \psi (x,t,\lambda) =0,
\end{aligned}
\end{equation}
with
\begin{equation}\label{eq:zz1}
\begin{aligned}
U(x,t, \lambda) &= \left( Q_0 + \lambda Q_1 - \lambda^2 J + \frac{1}{\lambda} \tilde{Q_1} - \frac{1}{\lambda^2} \tilde{J}\right), \\
V(x,t, \lambda) &= \Big( V_0 + \lambda V_1 + \lambda^2 V_2 + \lambda^3 V_3 - \lambda^4 K \\
&\quad + \frac{1}{\lambda} \tilde{V_1}+ \frac{1}{\lambda^2} \tilde{V_2}+ \frac{1}{\lambda^3} \tilde{V_3} -  \frac{1}{\lambda^4} \tilde{K}
\Big),
\end{aligned}
\end{equation}
where by ``tilde" we mean
\begin{equation}
\tilde{X} = -B X^T B^{-1}, \quad B =
\begin{pmatrix}
0 & -1 \\
1 & 0
\end{pmatrix}.
\end{equation}
Here $V_i, Q_i, J, K$ take values in the Lie algebra $\mathfrak{sl}(2)$. $V_i, Q_i$ are functions of $x$ and $t$ and $J$ and $K$ are constant matrices.
We will impose two reductions (types one and four from \eqref{eq:U-V}). Their effect on the potential of the Lax operator is given by
\begin{equation}\label{eq:GI-black1-2}
\begin{aligned}
&1) \quad U^{\dagger} (x,t, \lambda^*) = U(x,t,\lambda), \\
&2) \quad \tilde{U}\left(x, t, \frac{1}{\lambda}\right) = U(x,t, \lambda).
\end{aligned}
\end{equation}
The same holds for $V(x,t,\lambda)$.

The compatibility condition $[ L, M] = 0$ leads to the following set of recursion relations:
\begin{equation*}
\begin{aligned}
& \lambda^6 : \quad [J,K]=0, \\
&\lambda^5:  \quad [J,V_3 ]= [K,Q_2] , \\
& \lambda^4 : \quad [J,V_2]= [K,Q_0]+[Q_1,V_3], \\
&\lambda^3:  \quad   [J,V_1] = i \frac{\partial}{\partial x} V_3 + [Q_1,V_2] + [Q_0,V_3] - [\tilde{Q_1},K]  , \\
&\lambda^2:  \quad   [J,V_0]= i \frac{\partial}{\partial x} V_2  + [Q_0,V_2]+  [Q_1,V_1] + [\tilde{Q_1},V_3] 
\end{aligned}
\end{equation*}\begin{equation}\label{eq:GI-CC}
\begin{aligned}&\qquad +  [\tilde{J},K],  \\
&\lambda^1:  \quad   i \frac{\partial}{\partial t} Q_1=  i \frac{\partial}{\partial x} V_1 +  [Q_0,V_1] + [Q_1,V_0]+  [\tilde{Q_1},V_2]  \\
&\qquad  - [J, \tilde{V_1}]  - [\tilde{J},V_3],  \\
&\lambda^0:   \quad  i \frac{\partial}{\partial t} Q_0=  i \frac{\partial}{\partial x} V_0 + [Q_0, V_0] + [Q_1, \tilde{V_1}] + [\tilde{Q_1}, V_1] \\
&\qquad - [J, \tilde{V_2}] - [\tilde{J}, V_2].
\end{aligned}
\end{equation}
We will impose two additional restrictions. The first is that the corresponding RHP should have canonical normalization (this means that \eqref{eq:LaxUV} holds).
The second will be an additional reduction. Let $\xi(x, t, \lambda)$ be a FAS. Then
\begin{equation}
\label{eq:GI:black3}
\xi(x, t, -\lambda) = \xi^{-1}(x, t, \lambda).
\end{equation}
This, together with \eqref{eq:LaxUV}, the reductions \eqref{eq:GI-black1-2}, and the recursion relations \eqref{eq:GI-CC}  leads to the following form for the coefficients of $U(x, t, \lambda)$ and $V(x, t, \lambda)$
\begin{equation}
\begin{aligned}
Q_0 &= V_2 = \begin{pmatrix}- 2 q p & 0 \\ 0 & 2 q p \end{pmatrix}, \quad
V_3 = \begin{pmatrix} 0 & 2 q\\ -2 p& 0 \end{pmatrix}, \\
%
V_1 &= \begin{pmatrix} 0 & i \partial_x q +2q \\ i \partial_x p - 2p & 0 \end{pmatrix}, \\
V_0 &= \mbox{ diag} \Big(
2q^2 p^2 + i q \partial_x p - i p \partial_x q - 4 q p , \, - 2q^2 p^2 - i q \partial_x p + i p \partial_x q +4 q p \Big),
\end{aligned}
\end{equation}
where $p = q^*$.
The $\lambda^1$ terms in \eqref{eq:GI-CC} give the following equation
\begin{equation}
\label{eq:MainEQ}
i \frac{\partial q}{\partial t} +  \frac{1}{2} \frac{\partial^2 q}{\partial x^2}   + 2 i  q^2   \frac{\partial q^*}{\partial x} + 4 q |q|^4   - 8 q \left| q  \right|^2   + 4q =0.
\end{equation}
This is GI equation with an additional cubic nonlinearity (and a linear term). The $\lambda$-independent
term vanishes, provided that $q$ is a solution of \eqref{eq:MainEQ}.

\textcolor{black}{
This equation has appeared in the list of integrable NLS-type equations classified by \cite{MiShaYa2,MiShaYa1}.
Indeed, consider the equations (1.1) in \cite{MiShaYa1}:
\begin{equation}\label{eq:Mi1.1}\begin{split}
\frac{\partial u}{ \partial \tau } = \frac{\partial^2 u }{ \partial^2 x } + f(u,v,u_x,v_x), \qquad
-\frac{\partial v}{ \partial \tau } = \frac{\partial^2 v }{ \partial^2 x } + g(u,v,u_x,v_x),
\end{split}\end{equation}
with $v=u^*$, $g = f^*$ and fix up the function $f$ according to eq. (1.10)  in \cite{MiShaYa1}, i.e.:
\begin{equation}\label{eq:Mi1.10}\begin{split}
f(u,v,u_x,v_x) = 2 a u v u_x + b u^2 v_x + \frac{b(a-b)}{2} u^3 v^2 + c u^2 v.
\end{split}\end{equation}
Let us now put
\begin{equation}\label{eq:u-q}\begin{split}
\tau = \frac{i}{2} t, \qquad u(x,\tau) = q(x,t) e^{-4it} , \qquad a=0, \qquad b=4i, \qquad c=-16.
\end{split}\end{equation}
Then one can easily check that the first equation in (\ref{eq:Mi1.1}) coincides with (\ref{eq:MainEQ}),
while  the second equation in (\ref{eq:Mi1.1}) is obtained from (\ref{eq:MainEQ}) by complex conjugation.
Similar arguments show one that:
i) eq. (\ref{eq:MainEQ}) is equivalent to eq. (4.3.15) in \cite{51a} and, ii) eq. (24) in \cite{51b} is a vector generalization to
(\ref{eq:MainEQ}).
}

\subsection{Ferromagnet type equation }
Let us denote by a tilde a given algebra  automorphism. Let us take the Lax pair in the form
\begin{equation}\label{eq:HF11}\begin{split}
&\left (  \frac{\partial }{ \partial x }+ S_0 + \zeta S(x,t) + \frac{a}{\zeta} \tilde{S}\right) \psi (x,t,\zeta)=0, \qquad S_0=\tilde{S}_0, \\
&\left (  \frac{\partial }{ \partial t } + M_0  + \zeta W(x,t) +\frac{a}{\zeta}\tilde{W} + \zeta^2 S(x,t) +\frac{a^2}{\zeta^2}\tilde{S}\right) \psi (x,t,\zeta)=0, \\
& M_0=\tilde{M}_0.
\end{split}\end{equation}
where $S\in\mathfrak{g},$  $a$ is a constant and $\zeta$ is a spectral parameter \footnote{In this example the letters $S$, $X,$ $a,$ $\alpha$ etc are not related to any quantities from the previous (sub)sections. }. Moreover, we assume that $S^2=\openone$ and thus $S$ defines an involution. Furthermore, a symmetric space is defined as $$ \mathfrak{k} =\{ X, \qquad SXS=-X\}.$$

 The equations arising from the Lax pair are
\begin{equation}\label{eq:HF12}\begin{split}
&-S_x+[S,W]+[S_0,S]=0,\\
& -\tilde{S}_x+[\tilde{S},\tilde{W}]+[S_0,\tilde{S}]=0\\
&S_t-W_x+[S,M_0]+[S_0,W]+a[\tilde{S},S]=0,\\
&\tilde{S}_t-\tilde{ W}_x+[\tilde{S},M_0]+[S_0,\tilde{W}]+a[S,\tilde{S}]=0\\
& S_{0,t} - M_{0,x} +[S_0,M_0]+ a[\tilde{S}, W]+a[S,\tilde{W}]=0.
\end{split}\end{equation}
 Then it is easy to spot that $S_x, S_t$ and $W$ belong to the symmetric space $\mathfrak{k}$. Moreover,
$$ \frac{1}{4} \mathrm{ad}_{S}^2 = \openone \quad \mathrm{on} \quad \mathfrak{k}.$$
  From the first two equations $$W(x,t) =S_0+ \frac{1}{4} [S, S_x], \qquad \tilde{W}(x,t) = S_0+\frac{1}{4} [\tilde{S}, \tilde{S}_x].$$
 We can take for simplicity $S_0=0,$ then
\begin{equation} \label{M0x}
 M_{0,x}=\frac{a}{4}[\tilde{S}, [S,S_x]]+\frac{a}{4}[S, [\tilde{S},\tilde{S}_x]]
 \end{equation} and one can write formally
$$ M_{0}=\frac{a}{4}\int_{-\infty}^{x} \left([\tilde{S}, [S,S_x]]+[S, [\tilde{S},\tilde{S}_x]]\right) dx'.$$ So it seems that $M_0$ is in general nonlocal. The evolution equation becomes
$$S_t-\frac{1}{4}[S,S_{xx}]+[S,M_0]+a[\tilde{S},S]=0.$$


 In the particular case when $\tilde{S}=-S^T$ and $\lim_{|x|\to \infty}S=J$ is real and diagonal with $J^2= \openone  ,$ we clearly have $$\lim_{|x|\to \infty}\tilde{S}=-J$$ and the asymptotic Lax operator satisfies
$$ \left (  \frac{\partial }{ \partial x }+  \left(\zeta  - \frac{a}{\zeta}\right) J \right) \psi_a (x,t,\zeta)=0 $$ with
\begin{equation} \label{exp}
\psi_a (x,t,\zeta)=\exp\left( - \left(\zeta  - \frac{a}{\zeta}\right) J  x \right).
\end{equation}

In order to compare the spectrum with the spectrum of the Heisenberg ferromagnet model as it appears in most textbooks \cite{FaTa,GVY} we introduce a spectral parameter $\lambda= \zeta/i=-i\zeta.$  Then $$ \zeta  - \frac{a}{\zeta}=-i\left(\lambda +\frac{a}{\lambda} \right)$$ and writing $\lambda=|\lambda|e^{i \gamma}$ we obtain
$$\zeta  - \frac{a}{\zeta}=-i\left(|\lambda|e^{i\gamma }- \frac{a}{|\lambda|}e^{-i\gamma}\right)=\left(|\lambda|- \frac{a}{|\lambda|}\right) \sin \gamma - i \left(|\lambda|+ \frac{a}{|\lambda|}\right) \cos \gamma.$$
Thus the continuous spectrum is where the exponent in \eqref{exp} is oscillatory, that is $\sin \gamma =0$, comprising the horizontal axis of the $\lambda$-plane and the set $|\lambda| = \frac{a}{|\lambda|}$  or $|\lambda|^2=a$  which is the circle of radius $\sqrt{a}$ in the case when $a>0$, as shown on fig. \ref{fig:3}, the upper panel.

Let us now move to the $S \in su(2)$ case.  We take the following parametrization:
\begin{equation} \label{S}
\begin{split}
& S= \begin{pmatrix} S_3 & S_1-iS_2 \\  S_1 + i S_2  & -S_3  \end{pmatrix}, \qquad S= S^\dag \\
&\tilde{S}= -US^TU^{-1}.
\end{split}
\end{equation}
Since the above transformation is an involution, then
$$S= U(US^TU^{-1})^T U^{-1} = U U^{-T}S U^{T}U^{-1}.$$  Therefore $ U^{T}U^{-1}=\pm \openone.$  The most general form of $U$ is
$$ U= \begin{pmatrix} \mathfrak{a} & \mathfrak{b} \\  -\bar{\mathfrak{b}}  & \bar{\mathfrak{a}}  \end{pmatrix}, \qquad |\mathfrak{a}|^2+|\mathfrak{b}|^2=1.  $$   The case $ U^{T}U^{-1}=- \openone$ or $U^T=-U$ has only one nontrivial representative, $$ U= \begin{pmatrix} 0 & -1 \\  1  & 0  \end{pmatrix},$$ and does not produce a new nonlinear evolution equation. Therefore, parameterising $$\mathfrak{a}=e^{i\alpha}\cos \theta, \qquad \mathfrak{b}=e^{i\beta}\sin \theta $$   with  $U^T=U$ we have the most general form of $U$ compatible with the involution, which necessitates $b=i\sin \theta $ and
\begin{equation} \label{U}
U= \begin{pmatrix} e^{i\alpha}\cos \theta & i\sin \theta \\  i\sin \theta  & e^{-i\alpha}\cos \theta  \end{pmatrix},
\end{equation}
where $\alpha$ and $\theta$ are constant real parameters. In the special case when $\sin \theta=0,$ $\cos \theta=1$ we have a diagonal transformation matrix
$$ U= \begin{pmatrix} e^{i\alpha} & 0 \\ 0  & e^{-i\alpha}  \end{pmatrix}.$$
Introducing a vector form notation
$$ \underline{S }=(S_1, S_2, S_3)^T, \qquad S_1^2+S_2^2+S_3^2=1 $$ from \eqref{S}, \eqref{U} we obtain $$\tilde{\underline{S }}=A\cdot\underline{S } $$ where the matrix $A$ is explicitly given by
\begin{equation} \label{A}
A= \begin{pmatrix} -\sin^2 \theta -\cos^2 \theta \cos 2\alpha & \cos^2 \theta \sin 2 \alpha & -\sin 2 \theta \sin \alpha \\
\cos^2 \theta \sin 2 \alpha & \cos^2 \theta \cos 2 \alpha - \sin^2 \theta & -\sin 2 \theta \cos \alpha \\
-\sin 2 \theta \sin \alpha & -\sin 2 \theta \cos \alpha & -\cos 2 \theta \end{pmatrix}.
\end{equation}

The matrix $A$ has the following properties,
$$ A^2=\openone,\qquad  A=A^T, \qquad A^T=A^{-1}.$$
Moreover, the diagonalization of $A$ is
$$ A= V \text{diag}(1,-1,-1) V^{-1}, \qquad V^{-1}=V^T. $$
 Next, we use the correspondence between the commutator of $su(2)$ matrices and the cross-product of 3-vectors:
$$ [X,Y]=(2i)\mathcal{M}(\underline{X}\times\underline{Y}) $$  where
 $$ \mathcal{M}(\underline{X})= \begin{pmatrix} X_3 & X_1-iX_2 \\  X_1 + i X_2  & -X_3  \end{pmatrix} \equiv X.$$
From \eqref{M0x}
\begin{equation}
\begin{split}
&\partial_x M_0 =a\frac{(2i)^2}{4}\mathcal{M}\left(\underline{\tilde{S}}\times (\underline{S}\times \underline{S}_x)+\underline{S} \times ( \underline{\tilde{S}}\times \underline{\tilde{S}}_x)\right)  \\
&= -a\mathcal{M}\left( (\tilde{\underline{S}}\cdot \underline{S}_x) \underline{S} + (\underline{S} \cdot \tilde{\underline{S}}_x)\tilde{\underline{S}}  - (\underline{S}\cdot \tilde{\underline{S}})(\underline{S}+\tilde{\underline{S}})_x \right) \\
&=-a(F_x({S}+\tilde{{S}})-2F({S}+\tilde{{S}})_x)
\end{split}
\end{equation}
 where \begin{equation} \label{F} F=\frac{1}{2}A_{ij}S_i S_{i}\end{equation} is a scalar (summation is assumed). Indeed, since $A$ is symmetric,
 $$\tilde{\underline{S}}\cdot \underline{S}_x= A_{ij}S_j S_{i,x}=\frac{1}{2}(A_{ij}S_j S_{i})_x$$ and similarly for $(\underline{S} \cdot\tilde{\underline{S}}_x).$

In a vector form
$$ \partial _ x \underline{M_0} =-a(F_x(\underline{S}+\tilde{\underline{S}})-2F(\underline{S}+\tilde{\underline{S}})_x) $$
 Next, we represent the vectors $\underline{S}=V V^{-1}\underline{S}$ and $\tilde{\underline{S}}=A\underline{S}= V \Lambda V^{-1}\underline{S}$  where $\Lambda=\mathrm{diag}(1, -1,-1)$, hence
 $$\underline{S}+ \tilde{\underline{S}}= V\mathrm{diag}(2,0,0)V^{-1}\underline{S}.$$
 Introducing the orthogonal transformation $\underline{\Sigma} = V^{-1}\underline{ S} $ with
 $$\Sigma_1^2+ \Sigma_2^2 + \Sigma_3^2=1$$
 and using the fact that
 $$2F=\underline{ S}\cdot (A\underline{S}) = \underline{\Sigma} \cdot (\Lambda \underline{\Sigma}) = \Sigma_1^2- \Sigma_2^2 - \Sigma_3^2=-1+2\Sigma_1^2$$
 we further obtain
\begin{equation}
\begin{split}
&{\underline{M}}_{0,x}=-a V \left( F_x\begin{pmatrix} 2\Sigma_1  \\ 0 \\ 0  \end{pmatrix}-2F\begin{pmatrix} 2\Sigma_1  \\ 0 \\ 0  \end{pmatrix}_x \right)\\
&= -2a V \cdot \begin{pmatrix} F_x \Sigma_1 -2F\Sigma_{1,x} \\ 0 \\ 0  \end{pmatrix} =-2a V \cdot \begin{pmatrix} \Sigma_{1,x} \\ 0 \\ 0  \end{pmatrix}.
\end{split}
\end{equation}
 Thus $M_0$ in this case is local. Moreover,
\begin{equation}
\begin{split}
&\underline{M}_{0}=-2a V \cdot \begin{pmatrix} \Sigma_{1} \\ 0 \\ 0  \end{pmatrix}=-a V(\openone+\Lambda)\cdot \underline{\Sigma}= -a V(\openone+\Lambda) V^{-1}\underline{S}= -a(\underline{S}+\tilde{\underline{S}}).
\end{split}
\end{equation}
 Hence $M_0=-a(S+\tilde{S})$ and the equation becomes
$$S_t -\frac{1}{4}[S, S_{xx}] + 2a[\tilde{S}, S]=0 $$
or with a proper redefinition of the time variable (by a constant of $2i$), in a vector form
$$\underline{S}_t= \frac{1}{4}\underline{S} \times \underline{S}_{xx} + 2 a\underline{S }\times \tilde{\underline{S}}=\frac{1}{4}\underline{S} \times \underline{S}_{xx} +  2a \underline{S} \times (A\cdot \underline{S}).$$
The matrix $A$ is given in \eqref{A}. This equation is very similar to the integrable Landau-Lifshitz model \cite{FaTa}, in which the matrix $A$ however is diagonal and to the integrable Landau-Lifschitz-type model derived by A. Borovik in \cite{Bo} where the matrix $A$ is a projector of rank 1.
The Landau-Lifshitz type models describe, in general, the dynamics of an anisotropic ferromagnetic medium.

 The Hamiltonian of the model is
$$ H =\frac{1}{2} \int \left( \frac{1}{4}\underline{S}_x ^2 - 4a F(\underline{S}) \right) dx $$
where $F(\underline{S})$ is the quadratic form \eqref{F}. The Lie-Poisson bracket is
$$ \{F_1.F_2\}(S)= - \int \left\langle S,\left[\frac{\delta F_1}{\delta S},\frac{\delta F_2}{\delta S}\right ] \right\rangle dx = - \int \left\langle \left[S,\frac{\delta F_1}{\delta S}\right ],\frac{\delta F_2}{\delta S} \right\rangle  dx$$ where $\langle \cdot, \cdot \rangle $ is the pairing in the Lie algebra. Writing the Lie-Poisson bracket in a vector form where the pairing is the usual Euclidean scalar product
$$ \{F_1, F_2\}(S)=- \int \left\langle \underline{S}\times \frac{\delta F_1}{\delta \underline{S}} ,\frac{\delta F_2}{\delta \underline{S}}
\right\rangle dx,$$
(see the details in \cite{Ho1,Ho2}) gives
$$ \underline{S}_t=\{\underline{S}, H\} = \underline{S}\times \frac{\delta H }{\delta \underline{S} } .$$

Generalizations of ferromagnet-type models related to Lie-algebras and symmetric spaces are studied in \cite{YV,V2011,GGMV}.

\section{Discussions and conclusions}

The scope of the present paper is limited to the $\mathbb{Z}_h$ and  $\mathbb{D}_h$ reduction groups.
Of course there are quite a few examples of NLEE related to the tetrahedral $\mathbb{T}$, octahedral  $\mathbb{O}$ and even to the icosahedral  $\mathbb{Y}$
reduction groups, see \cite{Rhys} and the references therein. The deeper studies of these NLEE should be based on the relevant automorphic Lie algebras $\mathcal{A}(\mathfrak{g})$ \cite{LoMikh1,LoMikh2,LoSa,KniLoSa}.

\textcolor{black}{The fact that the generalized GI eq. (\ref{eq:MainEQ}) is integrable
has been known for long time now \cite{MiShaYa2,MiShaYa1}. The new facts about it
are the Lax representation which possesses $\mathbb{D}_4$ symmetry. The solution of the inverse scattering problem (\ref{eq:zz}), (\ref{eq:zz1})
requires construction of  its fundamental analytic solutions (FAS). Skipping the technical details we remark here, that these FAS
satisfy an equivalent RHP with nontrivial contour splitting the complex $\lambda$-plane into eight domains, see lower panel of fig. \ref{fig:3}.
The  calculation of the soliton solutions of (\ref{eq:MainEQ}) can be done via the Zakharov-Shabat dressing method.}

We point out also several open problems. \textcolor{black}{The list of examples of NLEE having $\mathbb{D}_h$ as  group of reductions can be naturally extended by considering more general Lax pairs than (\ref{eq:zz}), (\ref{eq:zz1}) related to
  symmetric spaces of higher rank. This could lead to Lax pairs for some of the vector NLEE whose integrability
  was proposed in \cite{51b}.   }
It is rather natural to expect that these NLEE possess hierarchies of Hamiltonian structures, whose
phase spaces are co-adjoint orbits of $\mathcal{A}(\mathfrak{g})$ passing through conveniently chosen element
$\mathcal{H}_p^{(k)} (\lambda)$.

More detailed studies of the mapping between the potential of $L$ and its scattering data would require the study of the Wronskian relations. While for $\mathbb{Z}_h$ and  $\mathbb{D}_h$ reduction groups such
construction seems to be rather straightforward, for  $\mathbb{T}$,  $\mathbb{O}$ and  $\mathbb{Y}$ this would require additional efforts.

\section*{Acknowledgments} R.I. is thankful to the organizers of the conference {\it New Trends in Applied Geometric Mechanics -- Celebrating Darryl Holm's 70th birthday} at ICMAT (Madrid, Spain) July 3--7, 2017 for their kind hospitality.
 One of us, V.S.G. is grateful to Prof. N. A. Kudrayshov  and to New Bulgarian University for their support.
R. I. I. and V. S. G. are grateful to Professors D. D. Holm and A. V. Mikhailov for many stimulating discussions. V. S. G and A. A. S. acknowledge funding from the NTS-Russia grant 02/101 from  23.10.2017.


\medskip
Received April 2018; revised March 2019.
\medskip

\end{document}